# Orbital angular momentum and energy loss characterization of plasmonic excitations in metallic nanostructures in TEM

*Matteo Zanfrognini [1,2], Enzo Rotunno[2*], Stefano Frabboni [1,2], Alicia Sit[3], Ebrahim Karimi[3], Ulrich Hohenester[4], and Vincenzo Grillo[2]*

1. Dipartimento FIM, Università di Modena e Reggio Emilia, via G.Campi 213/a, I-41125, Modena, Italy

2. CNR-NANO via G Campi 213/a, I-41125 Modena, Italy

3. Department of Physics, University of Ottawa, 150 Louis Pasteur, Ottawa, Ontario K1N 6N5, Canada

4. Institute of Physics, Karl-Franzens-Universitat Graz, Universitatsplatz 5, 8010 Graz, Austria

ABSTRACT. Recently, a new device to measure the Orbital Angular Momentum (OAM) electronic spectrum after elastic/inelastic scattering in a transmission electron microscope has been introduced. We modified the theoretical framework needed to describe conventional low loss electron energy loss spectroscopy (EELS) experiments in transmission electron microscopes(TEM) to study surface plasmons in metallic nanostructures, to allow for an OAM post selection and devise new experiments for the analysis of these excitations in nanostructures. We found that unprecedented information on the symmetries and on the chirality of the plasmonic modes can be retrieved even with limited OAM and energy resolutions.

Localized surface plasmon resonances are confined collective excitations of the conduction electrons in a metallic nanostructure, whose excitation energies depend both on the material and on the geometrical shape of the nanoparticle itself. In the last few years, the study of the properties of these peculiar resonances have become a topic of great interest in the context of physics, chemistry and material science. They have potential applications in a wide range of fields, from medicine[1] to SERS spectroscopy[2], and from optoelectronics to photovoltaics[3].

The primary technique for characterizing plasmon resonances is through exciting them with light in absorption and scattering experiments. The main limitations of this approach are:1) only the modes with a non-vanishing electric dipole (bright modes) can be excited and 2) the local fields associated to these excitations are not spatially mapped with enough resolution. A more flexible approach to their analysis is represented by electron energy loss spectroscopy (EELS) performed in a transmission electron microscope (TEM). EELS can spatially map, with sub-nanometre resolution, the fields associated to both bright and dark plasmonic resonances of a given metallic nanostructure[4][5][6].

In a typical EELS experiment, an electron probe with sub-nanometre transverse size is scanned on different points on the sample surface R, and the loss function $\Gamma(E, R)$ is evaluated at the excitation energy $E$ of a certain plasmonic resonance. As clearly shown by Boudarham and Kociak [7], this quantity turns out to be proportional to the square modulus of the electric field associated to this resonance, evaluated at point R and projected along the TEM optical axis. It can be seen immediately that this approach prevents the experimental measurement of the local phase (i.e. of the sign) of the fields characterizing these excitations; simultaneously, modes with energy separations smaller than the energetic resolution of the experimental apparatus are extremely difficult to be resolved.

A first attempt to have direct access to the phase of the excitations has been produced by Guzzinati and co-workers[8], who adopted a structured beam and collected only on-axis inelastically scattered electrons; in this way, only the energy loss due to the mode with spatial symmetry matching the one of the incoming wave can be detected. Possible limitations to this innovative approach could be:1) the requirement of using different structured beams (and, therefore, different measurements) to access the whole plasmonic spectrum of a nanostructure; 2) the necessity of preliminary knowledge of the spatial symmetries of the plasmon resonances to be probed (in order to choose an appropriate incoming electron wave) and 3) the fact that collecting only the inelastically scattered electrons along the TEM axis means decreasing the signal-to-noise ratio in the probed quantities [9].

It has been recently demonstrated experimentally the possibility of measuring the orbital angular momentum spectrum of a light [10] or an electron beam [11,12]: this type of measurement is possible by inserting two phase elements (OAM sorters) in the TEM column, performing the coordinate transformation from position to angular basis representation of the wave function[13]. Here, starting from the theoretical approach proposed in 8, we demonstrate the possibility of gaining information about the symmetries of the plasmonic fields by measuring not only the energy spectrum of the inelastically scattered electrons, but also their OAM spectrum, without using multiple structured beams as initial probes: so, differently from previous approaches we concentrate on measuring the final OAM state, keeping the initial state as constant. By setting up the electron microscope similarly to what explained in ref. 11 (see Supplementary information Figure S1 for more details about the electron optics configuration) it will be possible to disperse the OAM and energy loss spectra of the transmitted electrons in orthogonal directions, thus having access to the so called OAM resolved loss functions $\Gamma_\ell(E)$.

In the following, we will simulate the expected experimental results in a number of paradigmatic cases, while considering the effects of the finite resolutions both in energy and OAM due to an non ideal experimental set up. We demonstrate how such a double dispersed experiment could give access to novel information about plasmonic fields, with respect to conventional EELS measurements.

We provide now a brief description about how it is possible to simulate OAM resolved electron energy loss spectra. We start by describing the interaction between a probe electron in $\vec{r}$ and the charge density $\rho(\vec{r'})$ of the nanostructure through the following interaction Hamiltonian,

$$\widehat{H'} = -\int e \frac{\rho(\vec{r'})}{|\vec{r}-\vec{r'}|} d^3r' \qquad (1)$$

where the integration is performed over the nanoparticle's (NP) volume, and $\ell$ is the charge of the electron. For a weak interaction, the transition probability ($\Gamma$) from an initial electronic state $\Psi_i$ (with energy $E_0$) to a final state $\Psi_f^\ell$, with OAM value of $\ell\hbar$ is given using Fermi's golden rule as,

$$\Gamma \propto \sum_{f,n} \left| \int\int d\vec{r}d\vec{r'} \frac{\Psi_f^*(\vec{r})\Psi_i(\vec{r'})}{|\vec{r}-\vec{r'}|} <n|\rho(\vec{r'})|0> \right|^2 \delta(E_f - E_0 + \varepsilon_n - \varepsilon_0). \qquad (2)$$

Here, the OAM is defined along the probe electron's direction of propagation, i.e., the TEM optical axis (taken to be the z-axis), and $|0>$ and $|n>$ are, respectively, the ground state of the nanoparticle (with energy $\varepsilon_0$) and its $n^{th}$ excited state. Exploiting the definition of electronic susceptibility $\chi(\vec{r},\vec{r'},E)$ given in [14], and references therein, and considering only transitions for the probing electron to states with energy $E_f = E_0 - E$, it is possible to write,

$$\Gamma_\ell(E) \propto \sum_f \int\int d\vec{r}d\vec{r'} \Psi_f^\ell(\vec{r})\Psi^*_i(\vec{r}) Im(-W(\vec{r},\vec{r'},E))\Psi_f^{\ell*}(\vec{r'})\Psi_i(\vec{r'})\delta(E_f + E - E_0) \quad (3)$$

with $W(\vec{r},\vec{r'},E)$ being the screened interaction between an electron in $\vec{r}$ and one in $\vec{r'}$. Following the approach shown in 7, we can write such a function as,

$$W(\vec{r},\vec{r'},E) = \sum_m g_m(E)\phi_m(\vec{r})\phi_m(\vec{r'})^* \quad (4)$$

where $\phi_m(\vec{r})$ is the potential associated to the $m^{th}$ surface plasmon resonance of the metallic nanostructure, and $g_m(E)$ is the so called spectral function.

In the paraxial approximation, we take,

$$\Psi_{in}(x,y,z;\vec{K_{in}}) = e^{iK_{in}^z z}\psi_{in}(x,y) \quad (5.1)$$

$$\Psi_f^\ell(x,y,z;\vec{K_f}) = e^{iK_f^z z}\psi_f^\ell(x,y) \quad (5.2)$$

where $\psi_{in}(x,y)$ ($\psi_f^\ell(x,y)$) denotes the initial (final) electronic wave function in a plane perpendicular to the optical axis, while $\vec{K_{in}}$ and $\vec{K_f}$ are the electron wavevectors before and after the scattering process. This approach is correct every time $K_{in}^z$ and $K_f^z$ are much larger than the projection of the wavevector perpendicular to the optical axis.

By substitution of (4), (5.1) and (5.2) in Eq. (3), using non recoil approximation [14], summing over all the possible $K_f^z$ (exploiting the delta function in (3)) and performing the integration along z, we finally obtain the desired OAM resolved loss function,

$$\Gamma_\ell(E) \propto \sum_f \sum_m Im(-g_m(E)) \left| \iint dxdy \psi_f^{\ell}(x,y) \phi_m(x,y,q) \psi_i(x,y)^* \right|^2 \quad (6)$$

while $\phi_m(x,y,q)$ is the Fourier transform along z of the potential $\phi_m(\vec{r})$[15].

In the following calculations, we will assume $\psi_i(x,y)$ to be a Gaussian beam whose waist is comparable with the size of the plasmonic nanoparticles, while we will write the final electronic states $\psi_f^{\ell}(x,y)$ as [16][17],

$$\psi_f(x,y) = J_{|\ell|}(K_f r) e^{i\ell\varphi} \quad (7)$$

in which $J_{|\ell|}(K_f r)$ is a Bessel function of the first kind of order $\ell$, $K_f$ is the transverse wavevector (i.e. the projection of the electron wavevector $\vec{K_f}$ on the $xy$ plane, perpendicular to the optical axis), and $\ell$ is the winding number. In this way, the sum over the index $f$ appearing in Eq.(6) is now performed over an ensemble of such final states characterized by a fixed $\ell$ and with transverse wavevector $K_f$ in the interval $[0, K_{Max}]$, where $K_{Max}$ depends on the collection angle $\alpha$ of the detector as $\alpha = K_{Max}\lambda$, with $\lambda$ corresponding to the electron's De Broglie wavelength. In the following calculations, we have kept $K_{Max}$ fixed to 0.4 nm$^{-1}$ (semi-collection angles of about 1 mrad, see Supplementary Information Figure S2), while the electron energy $E_0$ has been taken equal to 300 KeV. Such an approach permits also to eventually include structured incoming electron beams, such as, for example, vortices and 2-lobed profiles as proposed respectively by refs. 18, 19 and 8. Our choice of using Gaussian beams with quite large beam waists is only associated with the possibility of exciting all the different plasmonic resonances of the nanostructure at the same time. In any case, our results are not dependent on the choice of using an effectively Gaussian electron beam profile; a plane wave-like illumination gives qualitatively the same results, changing only the relative intensities of the peaks related to different modes). The spectral function $g_m(E)$ depends on both the geometry and the dielectric properties of the considered metallic nanostructure, and its imaginary part (with a minus sign) is maximized at the $m^{th}$ surface plasmon resonance (SPR) energy. The analytical form of the imaginary part of this function is analogous to the one provided in ref. 7 by Boudarham and Kociak considering the metallic nanostructure embedded in a homogeneous medium with constant dielectric function and neglecting the contribution to the electron energy loss due to bulk excitations. Finally, $\phi_m(x,y,q)$ is the Fourier transform, along *z,* of the electric potential associated with the $m^{th}$ plasmonic mode; such a quantity does not depend on the dielectric properties of the material but only on the geometry of the considered sample.

Within this approach, the surface plasmon oscillations are treated classically, i.e. by computing the plasmon response function solving Maxwell's equations in the non retarded approximation through the boundary element method [14][15], implemented in MNPBEM toolbox [20]. The electron dynamics, however, are studied quantum mechanically using Fermi's golden rule, as outlined above. In the adopted theory, we neglect the contribution to the electron energy loss due to bulk plasmon excitations of the metal as their energies are expected to be larger than those of the localized surface plasmons of interest. We also suppose the nanostructure to be embedded in vacuum, neglecting in this way the effect of the substrate on which the structures are located. As reported in ref. 21, the effect of the substrate is both to red shift the SPR resonances and to increase the line widths of the plasmonic features, but the symmetry properties of the excitations, in which we are interested, are expected to be left unchanged by the substrate itself. As a final remark, we also point out that neglecting retardation effects should mainly give a blue-shift of the excitation energies [14], while leaving unchanged the symmetries of the modes which are the target of the proposed experiments. In any case, although in this work we have employed for simplicity the quasi- static approximation throughout, future work should also address the case of the full Maxwell's equations where a modal decomposition into resonance modes could be performed, in analogy to refs. 22 and 23.

The first example of OAM analysis of the inelastically scattered electrons by surface plasmons that we consider is the case of a nanostructure with cylindrical symmetry. We explore the didactic case of a nanodisk, but the following description can be extended to the case of any axially symmetric system like toroidal or spherical particles.

Because of the peculiar symmetry of this nanostructure, we expect the surface charge distributions associated to the different plasmonic modes to be characterized by an azimuthal dependence of the type $\cos(m\varphi)$ or $\sin(m\varphi)$, where m is a positive or null integer number. The same azimuthal behaviour is expected to be inherited by the associated electric eigenpotential $\phi_m(x, y, q) = \phi_m(r)\frac{e^{im\varphi} \pm e^{-im\varphi}}{2}$, where $\phi_m(r)$ describes the radial behaviour of the potential itself [24] (see Supplementary Information for a further clarification of this point, in particular Figure S4). Taking as incident state $\psi_i(r)$ a non-structured beam (e.g., a Gaussian beam with transverse dimensions at least equal to the nanoparticle diameter or a plane wave), the transition probability to a final state characterized by an OAM value of $\ell$ (i.e.$\Gamma_\ell(E)$), expressed in cylindrical coordinates, becomes:

$$\Gamma_\ell(E) \propto \sum_{K_f} \sum_m \left| \int dr \int_0^{2\pi} r J_{|\ell|}(K_f r) e^{i\ell\varphi} \phi_m(r) \left( e^{im\varphi} \pm e^{-im\varphi} \right) \psi_i(r)^* d\varphi \right|^2 \quad (8)$$

which gives us the selection rule,

$$\int_0^{2\pi} e^{i\varphi(\ell \pm m)} d\varphi \neq 0 \leftrightarrow \ell = \mp m. \quad (9)$$

Summarizing, each $\Gamma_\ell(E)$ is a function with peaks only at the excitation energies of the modes characterized by $m = \ell$; therefore, in an OAM resolved EELS experiment, one is able to distinguish the plasmonic resonances both according to their energy and their azimuthal symmetries with only a single measurement.

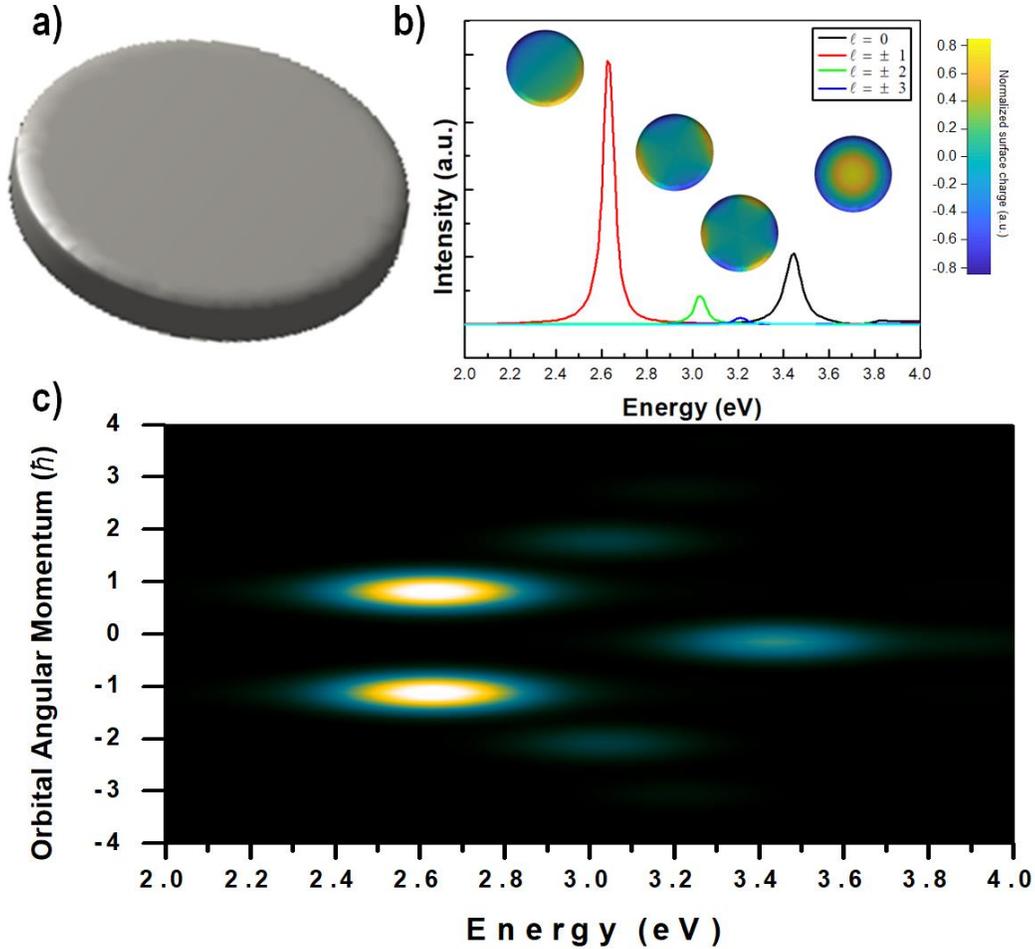

**Figure 1.** a) Tilted view of the Ag nanodisk. b) Simulated OAM-resolved EEL spectra for different OAM values (see legend). The surface charge distribution of each plasmonic mode is reported as an inset, where positive (negative) charge corresponds to blue (yellow) tone. c) 2D representation of the EEL spectra convoluted with a Gaussian function simulating the limited instrumental resolution ($\Delta E = 0.3 eV$ and $\Delta \ell = 0.5 \hbar$).

As an example, we performed numerical calculations for a silver nanodisk (silver dielectric function has been taken from Johnson[25]), with a height of 10 nm and a diameter of 70 nm (Figure 1a). The sample is illuminated by a Gaussian beam with a beam waist equal to 35 nm centered in the middle of the disk. We consider the behaviour of the loss function in the interval of energies [2;4] eV, where the first plasmonic modes locate in energy. The simulated OAM resolved EEL spectra, for different values of $\ell$, are reported in Figure 1b. As expected, for each OAM value, we have a peak at the energy of the plasmonic mode with the corresponding azimuthal symmetry. Therefore, the intense maximum observed in the spectrum obtained for $\ell = \pm 1$ corresponds to the dipolar edge mode (depicted in the inset), while the peaks found for $\Gamma_{\pm 2}$ and $\Gamma_{\pm 3}$ are respectively due to the quadrupolar and the hexapolar edge resonances of the nanodisk. As a last remark, the maximum observed for $\ell = 0$ is related to the breathing mode of the nanodisk [26], which is characterized by an eigenpotential not dependent on the variable $\varphi$.

Despite the results of the performed simulations fitting our precedent reasoning, from the experimental point of view, resolving the inelastically scattered electrons both in energy and in OAM is a difficult task. In particular, the limited instrumental resolution of the analysing system must be taken into account in order to obtain results which can be, in some way, compared with those found experimentally. In order to show the results closer to a realistic experiment, we have convolved the numerically obtained loss functions $\Gamma_\ell(E)$ with a Gaussian function keeping into account the broadening introduced by the experimental apparatus (see Supplementary Information Figure S3). The result is reported in Figure 1c where the values $\Delta E = 0.3 eV$ and $\Delta \ell = 0.5 \hbar$ are used for the energy and OAM resolution, respectively; these values are consistent with the resolution of modern state-of-the-art monochromated electron microscopes[14][24]. The simulation shows that, despite the experimental broadening, every loss function $\Gamma_\ell(E)$ is still distinctly peaked at the corresponding energies. This is an effect of experiments that are doubly dispersed in energy and OAM so that energy resolution by itself is not the main limiting factor.

We further noticed that, if the resolution is decreased to $\Delta E = 1 eV$ as for non-monochromatized Schottky field emission gun (FEG), thanks to the double OAM and energy loss dispersion, we are still able to capture the difference between the dipole and breathing modes.

As stated above, for systems with cylindrical (or more generally axial) symmetry, the surface plasmon modes are characterized by a well-defined azimuthal behaviour. If the symmetry is slightly broken, for example considering the structure shown in Figure 2a, the surface charge no longer follows the simple azimuthal distribution described above. In any case, it is reasonable to expect that, by destroying the cylindrical symmetry, the resulting modes can still be written as a superposition of the nanodisk modes[27]. We can thus expect that a given plasmonic resonance of the morphed structure can have a surface charge distribution (and so an eigenpotential) given by the sum of functions with a $\varphi$ dependence of the type $cos(m\varphi)$ and $sin(m\varphi)$, with different values of *m* for the same plasmonic mode (see Supplementary Information Figure S5 and Figure S6 for a complete decomposition of some of the resonances of the morphed structure). Taking into account what was explained before, we then expect to observe peaks at the same energy for loss functions computed for different $\ell$. More clearly, if at the excitation energy $E_\alpha$ of the mode α of the deformed disk we have a maximum for both $\Gamma_{\pm m_1}(E_\alpha)$ and $\Gamma_{\pm m_2}(E_\alpha)$, it means that such a resonance can be understood as given by the hybridization of the modes with azimuthal numbers $m_1$ and $m_2$ of the nanodisk[27]: such information cannot be gained using conventional EELS techniques.

We performed simulations for the structure shown in Figure 2a: a nanoellipse which has been obtained distorting the nanodisk in Figure 1a according to the approach described in the Supplementary Information (the major axis is equal to 84 nm, while the minor one is 60 nm). We assume an incoming beam analogous to the one adopted for the nanodisk.

Looking at the OAM-EEL resolved spectra reported in Figure 2b, we immediately notice that the two degenerate dipolar edge modes of the nanodisk split in energy and give very intense peaks for $\ell = \pm 1$. At the same time, it is simple to observe that the functions $\Gamma_{\pm 2}$ and $\Gamma_0$ have a common peak at an energy of 2.975 eV; however, $\Gamma_{\pm 2}$ has a maximum at 3.05 eV not observable in the trend of $\Gamma_0$. In good agreement with the observations reported in ref. 27, this effect can be justified by assuming that the two degenerate quadrupolar edge modes of the nanodisk separate in energy and the lower energy one mixes with the breathing mode: this common peak for $\Gamma_{\pm 2}$ and $\Gamma_0$ at the same energy can be considered as an experimental demonstration of this mode hybridization.

In Figure 2c, we present the expected results of a real life experiment, performing the same convolution procedure outlined in the case of the nanodisk. Even if the two close peaks for $\Gamma_{\pm 2}$ cannot be resolved as their separation is smaller than the assumed broadening in energy, it is immediate to notice that both the loss functions computed for $\ell = \pm 2$ and $\ell = 0$ have strong intensities at about 3.0 eV, signaling the mode mixing described before, even with a non-ideal experimental set up.

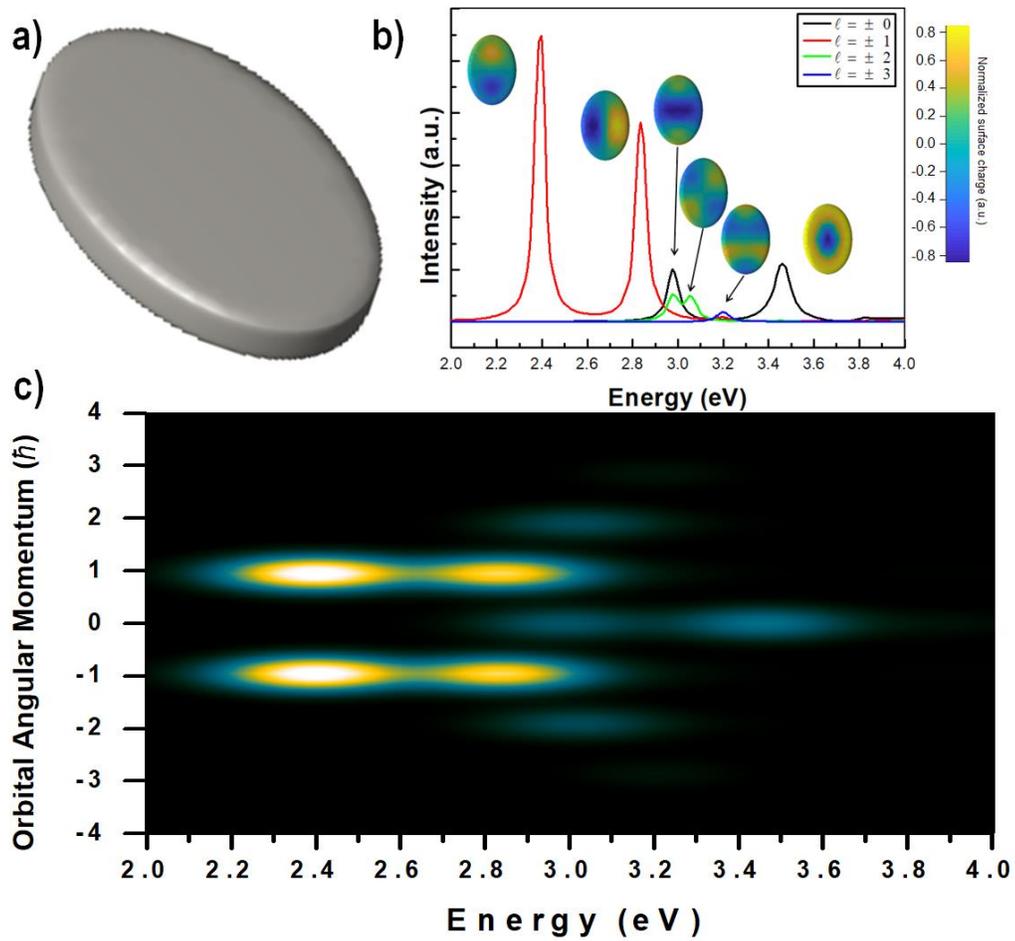

**Figure 2**: a) Tilted view of the Ag elliptical nanodisk. b) Simulated OAM-resolved EEL spectra for different OAM values (see legend). The surface charge distribution of each plasmonic mode is reported as insets. c) 2D representation of the EEL spectra convoluted with a Gaussian function simulating the limited instrumental resolution ($\Delta E = 0.3 eV$ and $\Delta \ell = 0.5\hbar$): notice a common strong intensity at about 3.0 eV for both $\ell = \pm 2$ and $\ell = 0$, pointing out the hybridization of the quadrupolar and the breathing modes of the nanodisk.

OAM-resolved EELS experiment are well suited for studying the azimuthal behaviour of plasmon resonances in non-cylindrically symmetric nanostructures, such as the nano-ellipse, but the discussion can be naturally extended to more complex, and technologically appealing, systems. This is the case of the plasmon resonances of cyclic formations of metal nanoparticles, known as metamolecules or plasmonic oligomers,[28][29][30] which are characterized by magnetic excitations.[31]

We consider here the case of the system reported in Figure 3a), composed of four identical elliptical nanodisks, each one having major axis, minor axis and height measuring 120 nm, 60 nm and 10 nm, respectively. The nanoparticles are disposed along the 160 nm long sides of a square resulting in a minimum distance between them of 21 nm, and they are illuminated with a Gaussian beam a with large beam waist (300 nm), centered in the middle of the square.

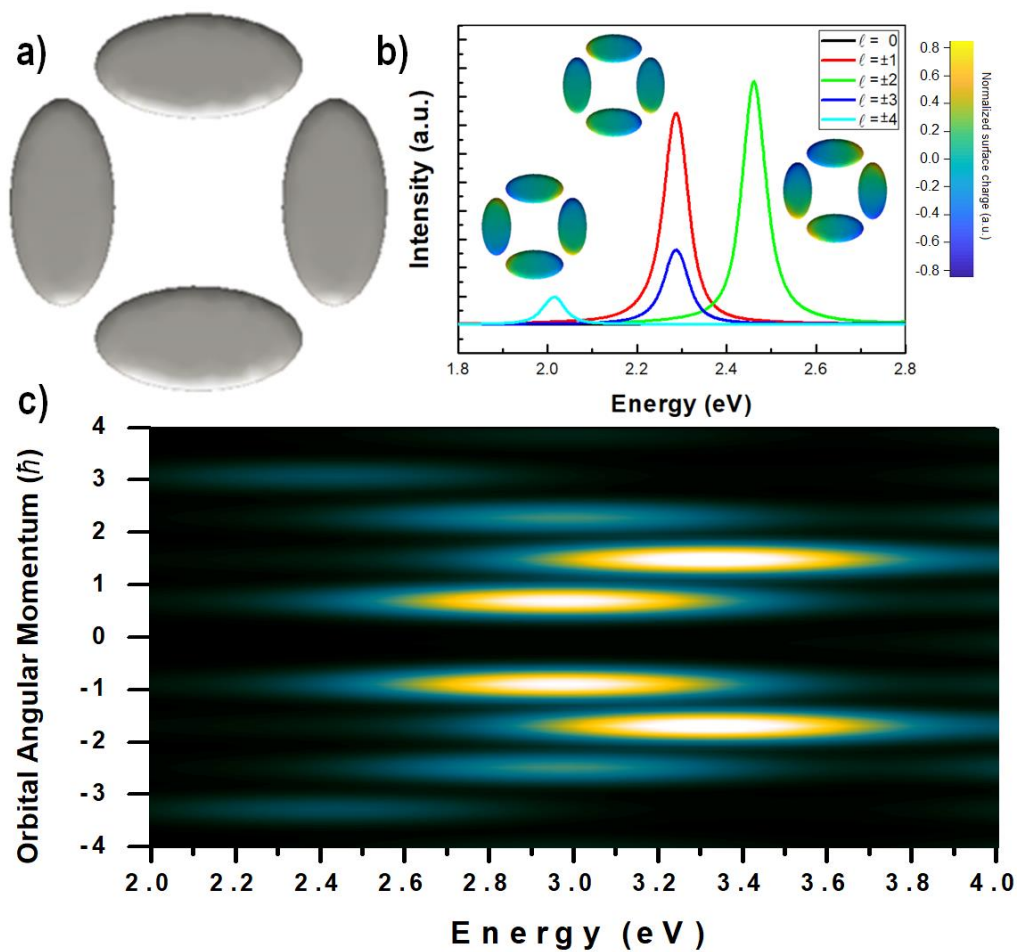

**Figure 3** a) Top view of the plasmonic molecule composed of four identical Ag elliptical nanodisk. b) Simulated OAM-resolved EEL spectra for different OAM values (see legend). The surface charge distribution of each plasmonic mode is reported as insets. c) 2D representation of the EEL spectra convoluted with a Gaussian function simulating the limited instrumental resolution ($\Delta E = 0.3 eV$ and $\Delta \ell = 0.5 \hbar$).

Figure 3b shows the calculated OAM-resolved EEL spectra in the energy range spanning from 1.8 to 2.8 eV, where the first three plasmonic modes, in order of energy, are located. Alongside each loss peak, reported as insets, are the respective surface charge distributions.

The lowest energy plasmon mode (2.02 eV) is characterized by a surface charge distribution (and so a potential) which changes its sign eight times in a complete path around the axis of the structure (see Supplementary information Figure S7 for further discussion).It is therefore reasonable to assume a potential with an in-plane azimuthal trend given by $\cos(4\varphi)$ or $\sin(4\varphi)$ which produce only peaks for $\Gamma_{\pm 4}$ at the energy of this mode. Analogously, the charge distribution associated to the resonance at 2.46 eV changes sign four times, so it is associated to a potential with $\cos(2\varphi)$ or $\sin(2\varphi)$ azimuthal profile which rationalizes the maxima only for $\Gamma_{\pm 2}$ at this energy.

More complex is, instead, the case of the plasmon peak at 2.28 eV as the charge distribution behaviour with $\varphi$ cannot be simply deduced graphically.A complete modal decomposition (see Supplementary Information Figure S7) suggests that the potential associated to this mode is characterized by very intense contributions of the type $e^{\pm i\varphi}$ and $e^{\pm i3\varphi}$, which give intense peaks for $\Gamma_{\pm 1}$ and $\Gamma_{\pm 3}$ at the energy of this resonance.

With this example, we point out the capability of deriving information about the symmetries and the signs of the potentials associated to plasmon resonances of unconventional nanostructures.

Such an example also underlines the possibility of exploiting such a technique to resolve plasmon resonances that are nearly degenerate in energy. For example, focusing on the two modes at 2.28 eV and 2.46 eV, even if their energetic separation is smaller than the broadening in energy of the incoming electron beam, their contributions to the electron losses can be separated as they provide peaks for different $\Gamma_\ell$. This is clearly noticeable once we look at the simulated experimental spectra shown in Figure 3c, where the peaks associated to these two modes are clearly distinguishable, despite the introduced broadening both in energy and OAM.

As a final application of OAM resolved EELS, we focus on the study of chiral plasmonic nanostructures. As already pointed out by Asenjo Garcia in ref. 32, chiral (i.e. not mirror symmetric) 3D plasmonic nanostructures are expected to produce electron OAM dichroism once illuminated by conventional beams, i.e. differences in intensities between the loss functions $\Gamma_{+\ell}(E)$ and $\Gamma_{-\ell}(E)$. In order to quantify the strength of such dichroic effects, we have followed [32] in defining the following figure of merit as,

$$D_{|\ell|}(E) = 100 \cdot \frac{\Gamma_{+\ell}(E) - \Gamma_{-\ell}(E)}{\Gamma_{+\ell}(E) + \Gamma_{-\ell}(E)}. \quad (10)$$

The chiral assemblies of metallic nanoparticles analysed in ref. 32 are expected to provide dichroic signals of the order of 10-15%. In the following, we present simulations performed for a metallic nanostructure which should exhibit giant electron dichroism, i.e values of $D_{|\ell|}(E)$ (for $|\ell| = 1$) of the order of 100%, which could greatly simplify the experimental detection of this effect (already attempted in ref. 33 for a cluster of nanospheres).

This structure in question, reported in Figure 4a, is made of two identical open silver nanorings, each of them characterized by a height of 10 nm, an external radius of 50 nm and an internal one of 25 nm; the spacing between the two nanorings is 20 nm and the opening angle has been fixed to 30° in all the calculations. The rings are rotated with respect to one another in order to break the mirror symmetry, as one can observe from the image. This structure may remind of a cylindrical conductor.[34] Such a structure was described in a recent paper in the context of interaction induced change of the OAM state. However this structure is deeply different due to the complete break of the symmetry.

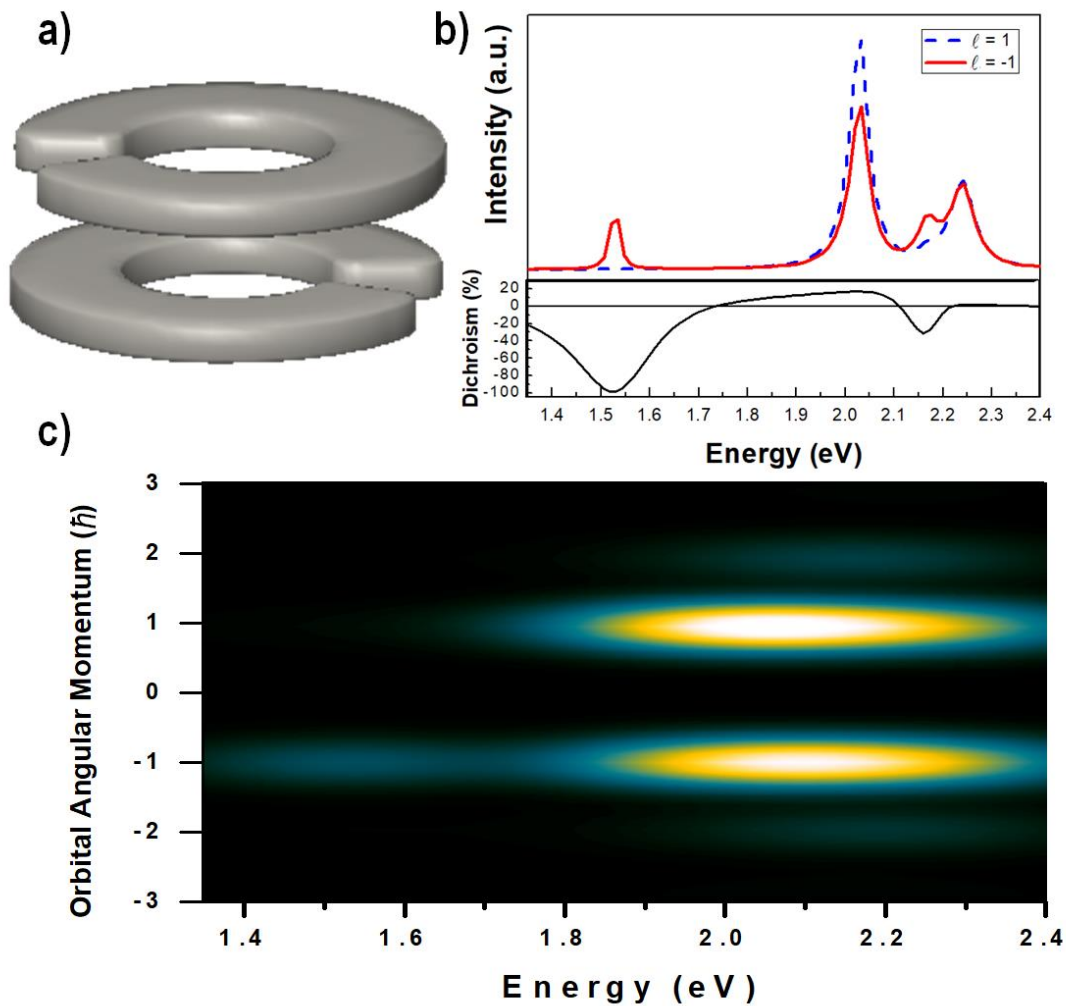

**Figure 4.** a) Tilted view of the chiral plasmonic structure. b) Simulated OAM-resolved EEL spectra for the $\ell = +1$ (blue dashed line) and $\ell = -1$ (red line) components: notice the presence of a peak only for $\ell = -1$ at 1.53 eV, which is responsible for a giant electron dichroism effect. c) 2D representation of the EEL spectra convoluted with a Gaussian function simulating the limited instrumental resolution ($\Delta E = 0.3 eV$ and $\Delta \ell = 0.5\hbar$). It is immediate to observe, despite the experimental broadening, a very pronounced asymmetry between the signals for $\ell = +1$ and $\ell = -1$.

In Figure 4b), we show the energy loss spectra for $\ell = 1$ and $\ell = -1$, the functions $\Gamma_{+1}(E)$ and $\Gamma_{-1}(E)$ respectively, calculated for an incident Gaussian beam with a beam waist of 50 nm, centered in the common center of the two rings. The comparison of the spectra immediately shows strong differences, in particular at the energy $E' = 1.53\ eV$ where the loss function $\Gamma_{-1}(E)$ is predominant while the opposite polarity $\Gamma_{+1}(E)$ is almost completely absent, leading to a dichroic signal $D_1(E)$ close to 100%, while a dichroic effect (even if much less pronounced) is also present in the peak at 2.17 eV. Conversely, the peak at $E'' = 2.03\ eV$ is more intense for the $\ell = 1$ component even though the difference is only $D_1(E'') = 20\%$ (see Supporting Information Figures S8 and S9 for a complete OAM decomposition of these modes). A realistic experiment, simulated in Figure 4c), is also able to observe the dichroism: by comparing the two spectra for $\ell = +1$ and $\ell = -1$, one can readily notice the complete absence of the peak at 1.53 eV in the $\ell = 1$ spectrum: notice that if this energy could be isolated, we would obtain a plasmon based generation of an electron vortex as in ref. 35 but without the use of light to excite the plasmons. The three peaks at 2.03, 2.17 and 2.24 eV cannot be individually resolved due to the limited resolution of the experimental setup; however, the maximum of the broad intensity distribution is shifted, reflecting the different intensity of the peaks. We recall that here we are not considering any post processing of the image that can improve the resolution and in principle retrieve the intensity of the single peaks.

In summary, we have demonstrated that, by combining the evaluation of the energy and OAM spectra of inelastically scattered electrons by a plasmonic nanostructure, it is possible to obtain additional information about the symmetries and also the chirality of the fields associated to these systems, by performing only a single measurement, i.e. without the need of modifying the features of the incoming electron wave. We proposed a possible measure to directly experimentally access the way in which plasmon modes mix together once a metallic nanostructure is distorted. We have also pointed out the possibility to exploit the double dispersion in OAM and energy loss to resolve peaks due to plasmon resonances which are separated in energy by a quantity smaller than the experimental resolution. All these information are not achievable exploiting techniques (e.g. conventional EELS in TEM or Photo Emission Electron Microscopy) normally used to map the electric fields of localized surface plasmon resonances of structures with characteristic sizes of few tens of nanometers. Furthermore, the use of the OAM sorter could be also very useful in the approach to the inverse problem of retrieving particle shapes by the explicit plasmon characteristics.

ASSOCIATED CONTENT

**Supporting Information.** Details about the set up needed to perform the proposed experiments and annotations to the performed simulations and to the interpretation of the presented simulated spectra.

AUTHOR INFORMATION

**Corresponding Author**

*E-mail: enzo.rotunno@nano.cnr.it

**Author Contributions**

The manuscript was written through contributions of all authors. All authors have given approval to the final version of the manuscript.

**Funding Sources**

This work is supported by Q-SORT, a project funded by the European Union's Horizon 2020 Research and Innovation Program under grant agreement No. 766970. A.S. and E.K. acknowledge the support of the Ontario Early Researcher Award (ERA) and the Canada Research Chair (CRC) program.